\documentclass{emulateapj}

\newcommand{\ntwo}{N$_{2}$H$^{+}$}
\newcommand{\ntwoi}{N$_{2}$H$^{+}$ 1$_{01}$--0$_{12}$}
\newcommand{\hcop}{HCO$^{+}$}
\newcommand{\meth}{CH$_{3}$OH}
\newcommand{\cs}{C$^{34}$S}
\newcommand{\htcop}{H$^{13}$CO$^{+}$}
\newcommand{\hcn}{H$^{13}$CN}
\newcommand{\iras}{IRAS~23033}
\newcommand{\msun}{M$_{\odot}$}
\newcommand{\um}{$\mu$m}
\newcommand{\kms}{km~s$^{-1}$}
\newcommand{\sig}{$\sigma$}
\newcommand{\cmt}{cm$^{-3}$}
\newcommand{\cmtwo}{cm$^{-2}$}

\newcommand{\mjb}{mJy~beam$^{-1}$}
\newcommand{\mjbkms}{mJy~beam$^{-1}$~km~s$^{-1}$}

\begin{document}

\title{Deconstructing the High-Mass Star-Forming Region IRAS 23033+5951}

\author{Michael A. Reid}
\affil{Harvard-Smithsonian Center for Astrophysics, Submillimeter Array 
Project, 645 North A'ohoku Pl., Hilo, HI, 96720, USA}

\author{Brenda C. Matthews} 
\affil{Herzberg Institute of Astrophysics, National Research Council of 
Canada, Victoria, BC V9E 2E7, Canada}

\begin{abstract}

        We report interferometric observations of the high-mass 
star-forming object IRAS 23033+5951.  Our observations reveal two massive 
molecular cloud cores, designated IRAS 23033+5951-MMS1 and IRAS 
23033+5951-MMS2.  MMS1 has already formed a massive protostar 
and MMS2 appears to be on the verge of doing so.  The latter 
core may be an example of a massive analogue to a ``Class 0'' star-forming 
object.  The more evolved core shows some evidence of \ntwo\ destruction
near the protostar, consistent with similar findings in low-mass 
star-forming objects.  In addition to the already-known prominent \hcop\ 
outflow, our SiO 2--1, and \meth\ 2--1 maps show evidence for two more 
candidate outflows, both presumably less powerful than the main one.  
Both cores are embedded in an elongated feature whose major axis is 
oriented almost exactly perpendicular to the axis of the most prominent 
outflow in the region.  Although it has many of the characteristics of a 
disk, the 87,000~AU (0.42~pc) diameter of this structure suggests that 
it is more likely to be the flattened, rotating remnant of the natal 
molecular cloud fragment from which the star-forming cores condensed.  We 
conclude that IRAS 
23033+5951 is an excellent example of massive star formation proceeding in 
relative isolation, perhaps by the method of monolithic collapse and disk 
accretion.

\end{abstract}


\section{Introduction}

	Two theories of massive star formation dominate the literature 
(see \citealt{mo07} for a recent comprehensive review).  
According to the first, massive stars form by the gravitational collapse 
of molecular cloud cores, which have masses of tens to hundreds of solar 
masses.  The molecular cloud cores are themselves produced by the 
fragmentation of molecular cloud clumps, which have masses of hundreds to 
thousands of solar masses.  In this ``monolithic collapse'' model, massive 
cores are envisioned to collapse to form single massive stars or small 
multiple systems via a process schematically similar to that which forms 
low-mass stars \citep{mt03,kmk05,k06}.  The other prominent theory of 
massive star formation, known as competitive accretion, holds that all 
stars begin as low-mass protostars which then compete to accrete some 
fraction of the mass of their parent molecular cloud clump, sometimes 
coalescing in the process \citep{bbz98,b01a,b01b,bvb04}.  In the 
competitive accretion scenario, massive stars are simply those which 
accrete most successfully.  It is hard to imagine that competitive 
accretion does not occur at some level in all clustered star-forming 
regions.  However, recent theoretical work suggests it is not the dominant 
mode by which massive stars form \citep{kmk05}.  Direct evidence capable 
of distinguishing these modes of star formation remains elusive.  To prove 
that massive stars can form by the collapse of individual molecular cloud 
cores would require several conditions to be verified by direct 
observation: (a) a massive protostar forming inside a well-defined 
molecular cloud core, (b) an infall signature in that molecular cloud core 
which is consistent with a high rate of protostellar accretion, (c) a 
massive circumstellar disk (whose properties are not yet well 
constrained), and (d) a collimated outflow with a high entrained mass.

	Relatively few massive star-forming objects with most or all of 
these characteristics have been identified.  Perhaps the best example is 
IRAS~20216+4104 \citep{ces97,ces99}, for which there is strong evidence of 
a collapsing Keplerian disk around a massive protostar.  Other examples, 
such as Cepheus~A~HW2 \citep{patel05}, are hotly disputed.  The 
principal obstacle in the study of massive protostellar systems is lack of 
spatial resolution.  Because massive protostars typically lie several 
kiloparsecs or more from the Sun, it is extremely challenging to spatially 
resolve their accretion disks, even with modern millimeter and 
submillimeter interferometers.  Our inadequate understanding of the 
chemistry of massive protostellar systems poses another significant 
challenge, making it difficult to select appropriate molecular transitions 
with which to study the kinematics of such systems.

	IRAS 23033+5951 (\iras\ hereafter) is an active star-forming 
region lying at a distance of 3.5~kpc in the Cepheus star-forming complex.  
Evidence of ongoing massive star formation has been detected in the form 
of compact radio continuum emission \citep{srid02,beu02b}, maser emission 
\citep{beu02b}, and two discrete dust continuum sources at 1.2~mm 
\citep{beu02}.  However, in contrast to many high-mass star-forming 
regions, relatively few discrete sources have been detected and there is 
evidence for only one substantial outflow \citep{beu02c,mz03}.  Thus, 
\iras\ provides an attractive opportunity to study massive star formation 
in a relatively quiescent, morphologically simple environment in which the 
usual complicating factors---crowding, abundant contamination from radio 
continuum emission, lack of spatial resolution---are mitigated.

	Our goal in studying \iras\ was to characterize its kinematics and 
determine whether it is a candidate example of a massive star-forming by 
monolithic collapse, as suggested by its simple morphology and known 
kinematics.  We observed \iras\ in several molecular transitions using the 
Berkeley-Illinois-Maryland Association (BIMA) array.  Each spectral line 
was selected to probe a particular kinematic aspect of \iras\, including 
outflows, rotation, infall, and turbulence.  In \S\ref{sec:obs}, we 
discuss the observations and the data reduction procedure.  In 
\S\ref{sec:morph}, we discuss the general morphology and evolutionary 
state of \iras, as determined from both the continuum and molecular line 
observations.  In \S\ref{sec:spec}, we concentrate on the kinematic 
results derived from the spectral line observations.  Our results are 
summarized in \S\ref{sec:summary}.

\section{Observations and Data Reduction}
\label{sec:obs}

\begin{deluxetable*}{llcccc}
\tablecaption{Molecular Lines Observed\label{tab:tunings}}
\tablewidth{0pt}
\tablehead{\colhead{Frequency} & \colhead{Molecular} & \colhead{Array} & \colhead{k$\lambda$} & \colhead{Velocity Res.} & \colhead{rms/channel}\\ \colhead{(GHz)} & \colhead{Line} & \colhead{Configurations} & \colhead{Range} & \colhead{(\kms)} & \colhead{(Jy beam$^{-1}$)} }
\startdata
86.243440 & SiO 2--1 & B, C & 2--69 & 0.337 & 0.13 \\
86.340167 & \hcn\ 1--0 & B, C, D & 2--68 & 0.085 & 0.21\\
86.754330 & \htcop\ 1--0 & B, C & 2--69 & 0.337 & 0.13 \\
89.188526 & \hcop\ 1--0 & C, D & 2--25 & 0.041 & 0.75 \\
93.173777 & \ntwo\ 1--0 & B, C & 2--74 & 0.079 & 0.079\\
96.412961 & \cs\ 2--1 & B, C & 2--74 & 1.21 & 0.058 \\
96.741377 & \meth\ 2--1 & B, C & 2--74 & 0.302 & 0.12\\
\enddata
\end{deluxetable*}

	Observations were made over 8 nights between 2003 October and 2004 
May using the BIMA interferometer in Hat Creek, CA.  All of the 
observations were made using single pointings toward the position $\alpha 
= 23^{\rm h}05^{\rm m}25\fs7$, $\delta = +60\degr08\arcmin08\arcsec$ 
(J2000).  Tracks were obtained in the B, C, and D configurations of the 
array and included tunings at 86, 89, and 93~GHz.  Not all of the spectral 
lines were observed with the same velocity resolution during every track.  
Table~\ref{tab:tunings} provides a complete summary of the lines observed, 
array configurations used, and the spectral line velocity resolutions used 
in most of our analysis.

	The data were flagged, calibrated, CLEANed and analyzed using the 
MIRIAD software package \citep{miriad}.  Observations of the nearby 
quasars 0102+584 and 2203+317 were made every 30 minutes during the 
observations and were used to correct instrumental and atmospheric phase 
and amplitude variations in the data.  Absolute flux calibration was 
performed using observations of Uranus taken during each track.  Based on 
the variations in the flux of the quasar from one epoch to the next, we 
estimate the accuracy of our flux calibration to be approximately 30\%.  
In order to maximize the signal-to-noise ratio of the data, we typically 
used natural weighting of the visibilities in generating our images. Where 
the analysis required higher spatial resolution than that provided by 
natural weighting, some of the data were processed using a MIRIAD 
``robust'' parameter of 0.5, at the cost of a lower signal-to-noise ratio.

	Our 3~mm continuum map was produced by combining continuum data 
from all of the different tunings around 3~mm.  The resulting CLEANed, 
naturally weighted image has a beam size of 5\farcs5$\times$4\farcs8 and 
an rms sensitivity of 2.9~\mjb.

\section{Morphology: Multiple Massive Cores}
\label{sec:morph}

\begin{figure*}
\begin{center}
\includegraphics[width=0.8\columnwidth,angle=270]{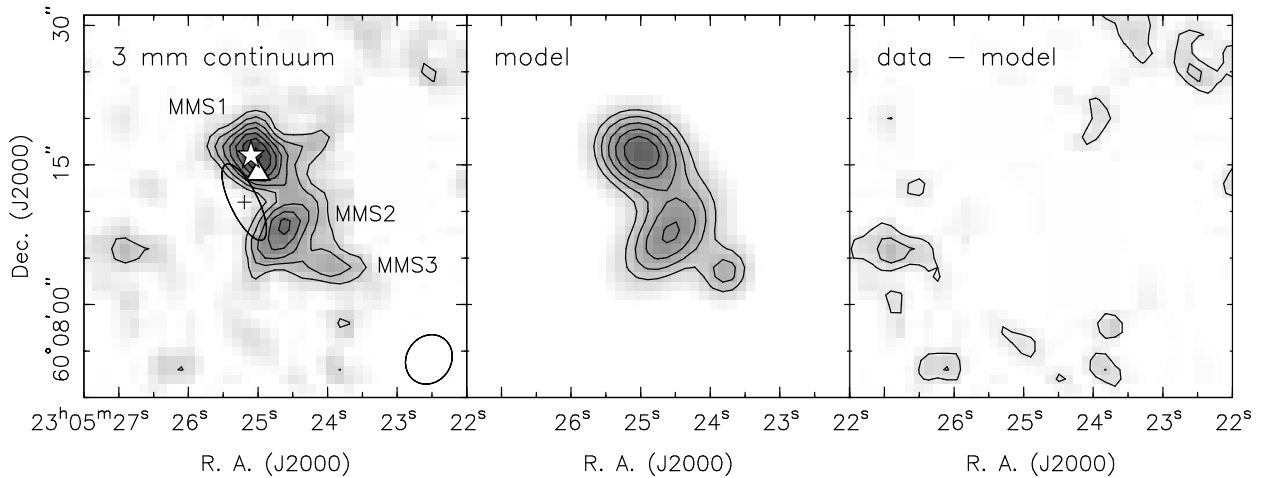}
\caption{Continuum emission at 3~mm (\emph{left}), a model of the 
continuum emission consisting of 2 elliptical gaussians and a point source 
(\emph{middle}), and the residuals obtained after subtracting the model 
from the data (\emph{right}).  The contours start at 3$\sigma$ in the two 
left panels and 2$\sigma$ in the right panel; the interval between 
contours is 1$\sigma$ in all three panels, where $\sigma = 2.9$~\mjb.
The 
ellipse in the lower right of the left panel shows 
the size and orientation of the synthesized beam.  
The star, triangle, and cross with error ellipse 
indicate the positions of the MSX source, H$_{2}$O 
maser \citep{beu02b}, and IRAS point source, 
respectively. \label{fig:contmodel}} 
\end{center} \end{figure*}

	The initial millimeter continuum observations of \iras\ with the 
IRAM 30~m telescope showed a singly peaked structure \citep{beu02}.  
Subsequent BIMA array observations of the source at 2.6~mm showed two 
continuum peaks \citep{beu02b}.  Our BIMA array 3~mm continuum image, 
shown in Figure~\ref{fig:contmodel}, reveals \iras\ to be composed of at 
least three continuum peaks joined by contiguous emission, plus a weak 
fourth peak visible to the east at the 3~\sig\ level.  We designate these 
three millimeter continuum peaks as IRAS~23033+5951-MMS1, 
IRAS~23033+5951-MMS2, and IRAS~23033+5951-MMS3 (hereafter, MMS1, MMS2, and 
MMS3) as shown in Figure~\ref{fig:contmodel}.  We used a simple model to 
determine the fraction of the continuum emission which is associated with 
each of the continuum peaks.  The model, which is shown in the middle 
panel of Figure~\ref{fig:contmodel}, fits the two bright peaks (MMS1 and 
MMS2) with an elliptical Gaussian and the third contiguous peak (MMS3) 
with a point source.  After model subtraction, the residuals in the region 
of the three main peaks are below 2~\sig, as shown in the right panel of 
Figure~\ref{fig:contmodel}.

	To compute the total masses of each of these continuum sources, 
we follow the prescription of \citet{h83}:
\begin{equation}
\label{eq:fluxtomass}
{\rm M} =
\frac{S_{\rm int}d^{2}}{\kappa B_{\lambda}({\rm T}_{\rm dust})}~~,
\end{equation}
\noindent where M, d, $S_{\rm int}$, $\kappa$, ${\rm T}_{\rm dust}$ are 
the 
mass, distance, 
integrated flux, dust opacity per unit mass column density, and dust 
temperature of the source, respectively.  $B_{\lambda}({\rm T}_{\rm 
dust})$ is the Planck function at wavelength $\lambda$ and temperature ${\rm T}_{\rm dust}$.  We assume ${\rm T}_{\rm dust}=30$~K, $d=3.5$~kpc 
\citep{harju}, a dust emissivity index of
 $\beta =1.5$, 
$\kappa = 0.1(250.0~\mu{\rm m}/\lambda)^{\beta}$~g$^{-1}$~cm$^{2}$ 
\citep{h83}, and a gas-to-dust mass ratio of 100.  Under these 
assumptions, we derive total masses of 225~\msun\ for MMS1, 205~\msun\ for 
MMS2, and 51~\msun\ for MMS3.  

	\citet{wil04} observed \iras\ using the JCMT and measured 850~\um\ 
and 450~\um\ integrated fluxes of 10.4$\pm$0.2~Jy and 48.6$\pm$2.6~Jy, 
respectively, for a region encompassing MMS1, MMS2, and MMS3.  Using these 
fluxes and assuming $\beta = 0.9$ and T$_{\rm dust}$ = 52~K, \citet{wil04} 
compute masses of 237~\msun\ at 850~\um\ and 106~\msun\ at 450~\um\ for 
the \iras\ complex.  Similarly, using single-dish measurements at 1.2~mm, 
\citet{beu02} derived a mass of 2327~\msun\ for \iras, assuming $\beta =2$ 
and T$_{\rm dust}$ = 52~K.  The differences between the calculated masses 
are caused by three factors: the differences in the measured fluxes, the 
differences in the assumed dust opacities, and the differences in the 
assumed dust temperatures.  Our interferometric maps filter out much of 
the flux visible in the single-dish maps.  Because our measurements were 
taken at a different wavelength than any of the single-dish measurements, 
the computation of the exact amount of flux filtered out by the 
interferometer is dependent on the assumed dust opacity and temperature.

	If we adopt the same flux-to-mass conversion prescription as 
\citet{wil04} (namely, assuming T$_{\rm dust}$ = 52~K and using the 
\citet{oss94} dust opacities extrapolated to 3.3~mm) we calculate masses 
of 131~\msun, 120~\msun, and 29.5~\msun for MMS1, MMS2, and MMS3, 
respectively.  We note, however, that the \citet{oss94} opacities were not 
originally tabulated for wavelengths as long as 3.3~mm, nor were they 
specifically meant to apply to massive star-forming regions.  Similarly, 
if we use the prescription of \citet{beu02} for converting integrated flux 
to mass, we find masses of 459~\msun, 419~\msun, and 103~\msun\ for MMS1, 
MMS2, and MMS3, respectively.  Clearly the mass uncertainties are quite 
large.  Nevertheless, even allowing for generous uncertainties in the 
flux-to-mass conversion and in the star formation efficiency, at least 
MMS1 and MMS2 appear to have enough mass to each form one or more massive 
stars.

	Gravitationally bound cores capable of forming massive stars 
typically have very high densities, $n \geq 10^5$~\cmt, and tend to have 
virial ratios, $\alpha_{\rm vir} = M_{\rm vir}/M$, close to unity 
\citep{p97,mt03}.  Do either of the spatially resolved cores in \iras\ 
meet these criteria?  We can compute the mean densities of the two 
brightest cores using their total masses (assumed to be 
225~\msun\ for MMS1 and 205~\msun\ for MMS2, as computed above) and their 
mean radii, defined as half of the mean of their major and minor axes, as 
determined from the fits shown in Fig~\ref{fig:contmodel}.  These mean 
radii are $0.14\pm0.02$~pc for MMS1 and $0.15\pm0.03$~pc for MMS2.  In 
this way, we compute mean gas densities of ($3\pm1$)~$\times~10^6$~\cmt\ 
and ($2\pm1$)~$\times~10^6$~\cmt, for MMS1 and MMS2, respectively.  Such 
high gas densities again qualify both cores as potential sites of massive 
star formation.
	
\begin{figure}
\includegraphics[width=0.8\columnwidth,angle=270]{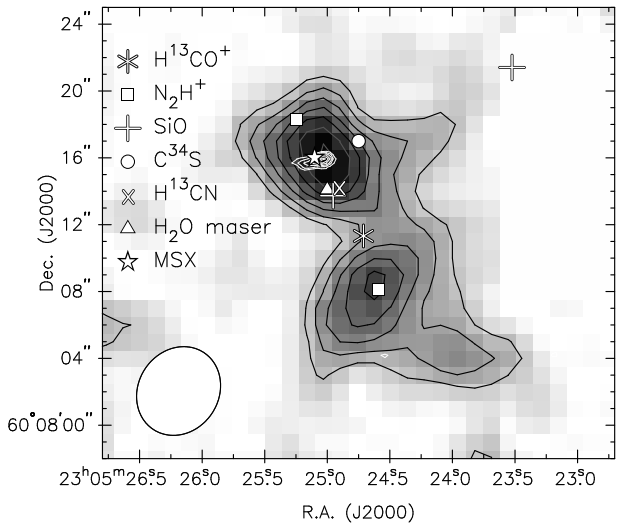}
\caption{Molecular line emission peaks (\emph{symbols}) overlaid on the 
3~mm continuum map (\emph{greyscale and contours}).  The meaning of each 
symbol is indicated in the legend.  For simplicity, the 
positional error bars are not shown; they are typically 
$\pm$1--2\arcsec\ in each coordinate.  The thin white contours underlying 
the MSX point source are the VLA 3.6~cm radio continuum contours from 
\citet{beu02b}, which have 0.7\arcsec\ resolution.  The coincidence 
of the continuum and \ntwo peaks in MMS2 suggests it is cold, 
and may be a pre-stellar massive star-forming core. 
The open ellipse in the lower left corner shows 
the size and orientation of the synthesized beam of the 3~mm continuum 
data.  
\label{fig:peaks}} 
\end{figure}

	Figure~\ref{fig:disklines} shows integrated intensity maps of 
\iras\ in four molecular lines which principally trace dense gas: \ntwo\ 
1--0, \htcop\ 1--0, \hcn\ 1--0, and \cs\ 2--1.  These four dense gas 
tracers peak at different locations.  The best column density tracers 
should be \ntwo\ 1--0 and dust, because both of these should be optically 
thin and relatively (although perhaps not entirely) unaffected by the 
complex carbon chemistry which can change the abundances of \htcop\ and 
\hcn\.  There is no indication in the \ntwoi\ line shape that this 
transition is optically thick.  Indeed, comparison of 
Figure~\ref{fig:contmodel} and Figure~\ref{fig:disklines} shows that the 
spatial distribution of \ntwo\ is very similar to the distribution of 
dust.  The peak positions of \ntwo\ and dust are separated by less than a 
beam width in both MMS1 and MMS2 (see also Fig.~\ref{fig:peaks}).  Hence, 
we conclude that, as expected, \ntwo\ 1--0 and dust both trace the densest 
gas in the region.  We can therefore use the \ntwoi\ line widths and the 
masses determined from the continuum emission to compute the virial ratios 
for MMS1 and MMS2.  We derive the \ntwo\ line widths from the isolated 
(i.e. non-blended) hyperfine component of the \ntwo\ line.  The \ntwoi\ 
line widths (full-width, half-power) in MMS1 and MMS2 are $2.5\pm0.3$ and 
$2.4\pm0.2$ \kms, respectively.  If we approximate the cores as spheres 
with uniform densities equal to the mean densities stated above and assume 
that the shape of the \ntwoi\ line is approximately Gaussian, then the 
virial masses of the cores are given by

\begin{equation}
M_{\rm vir} = \frac{5R(\Delta v)^{2}}{8{\rm G}\ln{2}}~~,
\label{eq:mvir}
\end{equation}

\noindent where $R$ is the radius of the core and $\Delta v$ is the 
observed \ntwo\ line width.  By this prescription, the virial masses of 
MMS1 and MMS2 are 178 and 184 \msun\ respectively.  Hence, their virial 
ratios are 0.79 and 0.90.  The virial parameter can be interpreted as a 
measure of a cloud's degree of gravitational binding, with values less 
than 1 indicating a gravitationally bound cloud.  Therefore, subject to 
the uncertainties in determining both the virial and dust continuum masses 
of the \iras\ cores, we conclude that they are both gravitationally bound.  
This interpretation is consistent with the fact that MMS1 is clearly 
forming at least one star.  It also suggests that MMS2 might already be 
forming a star, or may eventually do so.

\begin{figure*} 
\begin{center} 
\includegraphics[angle=270]{f2.eps} 
\caption{Integrated intensity of four lines in \iras.  Counter-clockwise 
from the top left panel, the lines are: \ntwo, \htcop, \cs, and \hcn.  The 
contours start at 3\sig\ in all panels 
and increase by steps of 4, 1, 1, and 1\sig, respectively.  The \ntwo\ 
intensity was integrated over all seven hyperfine components of the 
line.  The sizes and 
orientations of the four beams are indicated by the ellipses in the lower 
right corners of each panel.  White crosses indicate the positions of the 
3~mm continuum peaks from Fig.~\ref{fig:contmodel}.  The 1~\sig\ rms in 
the panels are, from left to right, 290~\mjbkms, 210~\mjbkms, 150~\mjbkms, 
and 320~\mjbkms.	
\label{fig:disklines}}
\end{center}
\end{figure*}

	Based on C$^{18}$O observations, \citet{on96} derived a mean 
virial parameter of $\sim$0.6 for star-forming molecular cloud cores in 
Taurus.  Similarly, \citet{tach} derive a mean virial ratio of 0.9 for 
C$^{18}$O cores in $\rho$~Oph, with smaller ratios being associated with 
more actively star-forming cores.  The C$^{18}$O cores in Taurus and 
$\rho$~Oph have typical masses in the tens of solar masses and radii of 
$\sim$0.2~pc.  Despite the fact that these cores may have sufficient mass 
to form high-mass stars, the evidence from their environments is that they 
will instead form clusters of low-mass stars.  The principal 
distinguishing factor between the higher-mass cores in $\rho$~Oph and 
those in \iras\ seems to be that the \iras\ cores have much higher surface 
densities: $N_{\rm H_{2}} \sim 10^{21.5}$~\cmtwo\ in $\rho$~Oph 
\citep{tach} vs. $N_{\rm H_{2}} \sim 10^{24}$~\cmtwo\ in \iras.  The high 
surface densities of the \iras\ cores are typical of those found in the 
massive star-forming clumps surveyed in CS and C$^{34}$S by \citet{p97}.  
Thus, very dense gas seems to be key requirement for the formation of a 
massive star, as predicted by both the monolithic collapse and competitive 
accretion scenarios.

\subsection{Chemical Signatures of Evolutionary States}
\label{sec:chemistry}
	
	The rapid evolution of the physical conditions in massive 
star-forming objects drives substantial chemical evolution in these 
objects.  The time evolution of the abundances of particular 
molecules has not yet been well established for massive cores, but we can 
take some cues from the chemical studies of low-mass star-forming cores.  
In particular, we note that numerous studies have shown that, as a 
collapsing low-mass core heats up, CO desorption from grain mantles causes 
substantial destruction of \ntwo.  The destruction of \ntwo\ lowers the 
\ntwo\ abundance near the center of the core (e.g. 
\citealt{lee04,dif04,ber02}).  This drop in the \ntwo\ abundance has been 
observed in the centres of several low-mass cores, such as L483 
\citep{jor04} and Barnard 1c \citep{mat06}.  \citet{pir07} have shown 
that, in massive star-forming objects containing IRAS point sources, there 
is a measurable decrease in the \ntwo\ abundance toward the column density 
peaks, again likely indicative of \ntwo\ destruction in these evolved 
sources.

	Although the sequence of chemical processes at work in high-mass 
star formation likely differs from that in low-mass star formation, the 
destruction of \ntwo\ by CO is likely to occur in both instances.  In 
\iras, the displacement between the dust and \ntwo\ emission peaks in 
MMS1, although small, suggests that the destruction of \ntwo\ near the 
massive protostar in MMS1 may already be underway.
Figure~\ref{fig:peaks} 
shows our 3~mm continuum image overlaid with the peak positions of five of 
the molecular species we observed, as well as the positions of the 
H$_{2}$O maser (taken from Table~2 of \citealt{beu02b}), the MSX point 
source, and contours of 3.6~cm VLA radio continuum 
emission\footnote{The position of the MSX point source was taken from the 
MSX point source catalog \citep{msxcat}.  The 3.6~cm radio continuum data 
were provided by H. Beuther.}.  The coincidence of the radio continuum and 
MSX emission with the dust peak in MMS1 suggests that a massive protostar 
is forming there.  The displacement between the dust and \ntwo\ peaks 
suggests that heating by the massive protostar may have caused the 
desorption of CO from dust grain mantles and the subsequent destruction of 
\ntwo\ near the MMS1 peak.  We emphasize, however, that the displacement 
between the \ntwo\ and dust peaks in MMS1 is less than a beam width, and 
may not be significant.  It is merely suggestive of chemical evolution in 
MMS1.

	Similarly, we can interpret the close correspondence of the dust 
and \ntwo\ peaks in MMS2 as evidence that it is cold, chemically 
unevolved, and presumably therefore also very young.  Its mass and 
apparent lack of chemical evolution suggests that MMS2 is a good candidate 
pre-stellar massive star-forming object.

	Our interpretation of MMS1 as chemically and dynamically more 
evolved than MMS2 is supported by the virial ratios of the two cores: 
lower in MMS1, indicating more gravitational binding, and higher in MMS2, 
indicating less binding and therefore perhaps an earlier evolutionary 
state (see \S~\ref{sec:morph}).

\section{Spectral Line Results: Kinematics}
\label{sec:spec}

\subsection{Outflows}

Several previous studies have demonstrated the presence of at least one 
outflow in \iras.  \citet{mz03} mapped the region in CO 3--2 using the 
10~m Heinrich Hertz Telescope telescope and found overlapping red- and 
blue-shifted outflow lobes emanating from a point not precisely coincident 
with the IRAS point source.  Using BIMA array CO 2--1 observations, 
\citet{b04} estimated the mass entrained in this outflow to be 119~\msun.  
Among the lines we observed, three are known outflow tracers: \hcop\ 1--0, 
\meth\ 2--1, and SiO 2--1.  These three lines give strikingly different 
pictures of the outflow activity in the \iras\ region.

\subsubsection{\hcop and SiO}
\label{sec:hcopsio}

\begin{figure}
\begin{center}
\includegraphics[width=0.8\columnwidth,angle=270]{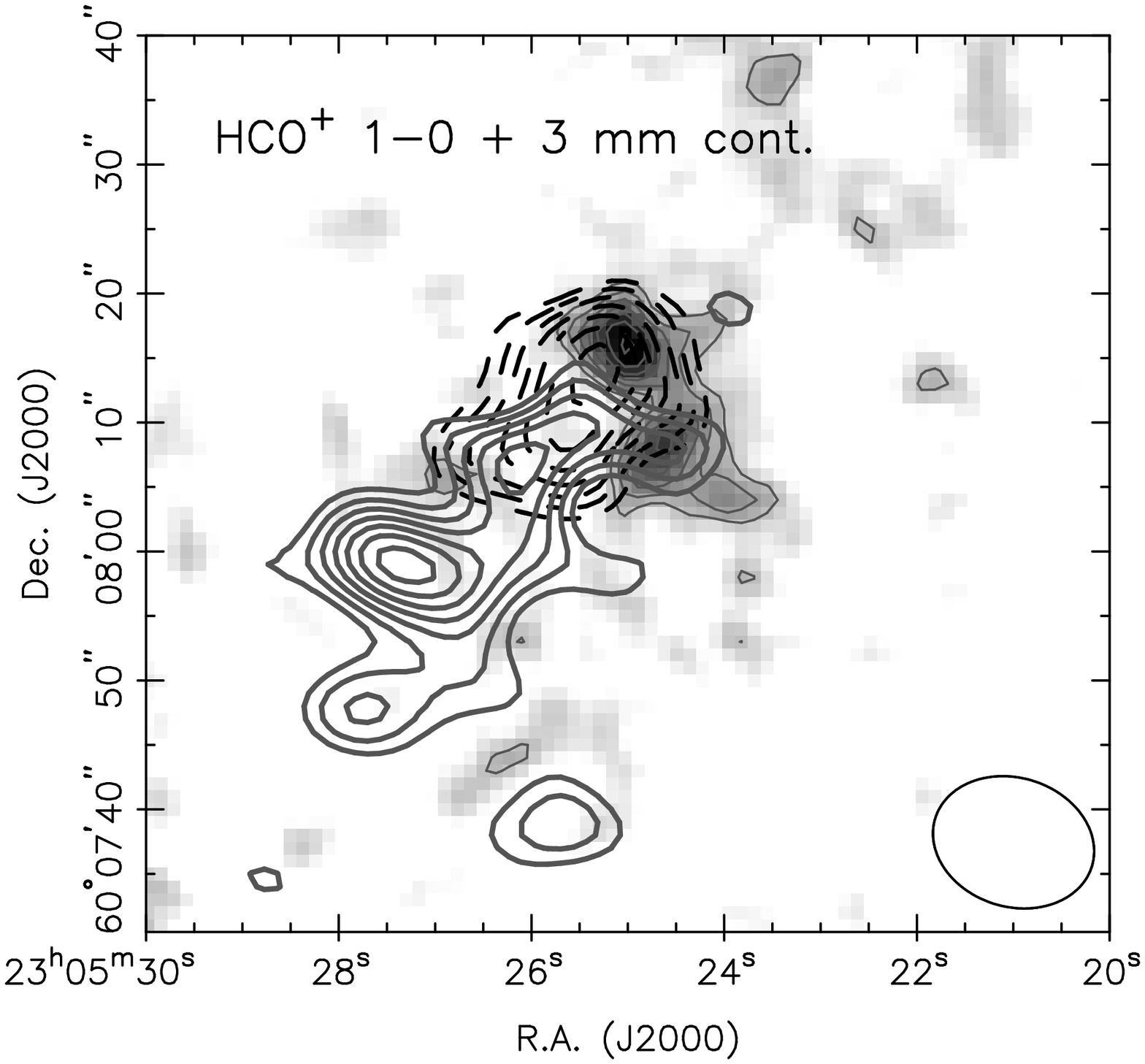}
\caption{Integrated intensity of high-velocity \hcop\ 1--0 emission 
plotted over 
continuum emission (\emph{greyscale and thin solid contours}).  The blue 
outflow lobe (\emph{dotted contours}) is integrated over the velocity 
range $-61.2$ to $-63.3$ \kms\ and the red lobe (\emph{thick solid 
contours}) 
is integrated over the range $-44.8$ to $-46.9$ \kms.  Contours begin at 
6~\sig\ for the red lobe and 4~\sig\ for the blue lobe, where \sig\ = 
108~\mjbkms, and increase by steps of 1~\sig. The ellipse in the 
lower right shows the size and orientation of the \hcop\ beam.  The 
blue-lobe \hcop\ emission prominently overlaps MMS1, though their peaks do 
not coincide.  The overlap of the red and blue 
lobes is probably an indication that the outflow is strongly inclined to 
the plane 
of the sky.  \label{fig:hcop_outflow}}
\end{center} 
\end{figure}

\begin{figure}
\begin{center}
\includegraphics[width=0.8\columnwidth,angle=270]{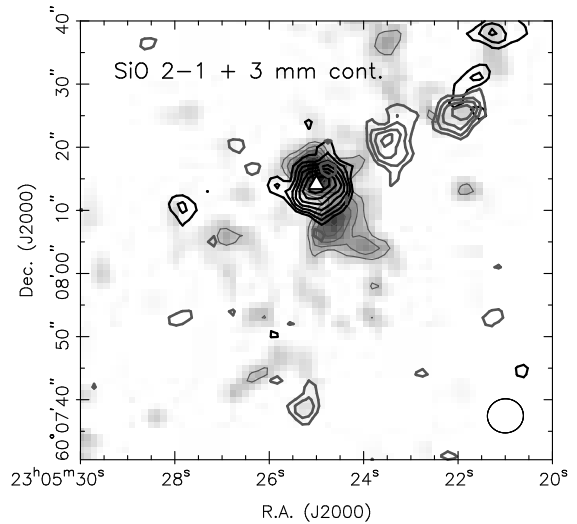}
\caption{Integrated intensity of SiO~2--1 plotted over 3~mm continuum 
emission.  The thick grey contours show the intensity of SiO 2--1 
integrated over the same velocity range as the red lobe of the \hcop\ 1--0 
outflow, namely $-44.8$ to $-46.9$ \kms\ (see 
Fig.~\ref{fig:hcop_outflow}).  
The black contours show the intensity of SiO 2--1 around the 
systematic velocity, integrated from $-51$ to $-55$ \kms.  No SiO emission 
was detected over the velocity range corresponding to the blue lobe of 
the \hcop\ outflow.  The thin grey contours and grey scale represent the 
continuum as in Fig.~\ref{fig:hcop_outflow}.  All contours start at 3\sig\ and increase by 
steps of 1\sig, where \sig\ = 120~\mjbkms\ for the red lobe and \sig\ = 
150~\mjbkms\ for the emission integrated around the systematic velocity.  
The 
white triangle indicates the 
position of the water 
maser, which is coincident with the local peak of the SiO emission.  The 
ellipse at the lower right 
shows the size and orientation of the SiO beam. \label{fig:sio_outflow}}
\end{center} 
\end{figure}

	Figure~\ref{fig:hcop_outflow} shows contours of red- and 
blue-shifted \hcop\ from our BIMA array observations superimposed on the 
3~mm continuum map.  This outflow clearly corresponds to the one seen by 
previous authors \citep{mz03,b04}.  The \hcop\ outflow is difficult to 
image because there is a significant background of \hcop\ associated with 
the underlying molecular cloud core, a source of confusion possibly 
amplified by the limited spatial resolution of our \hcop\ observations.  
In constructing Figure~\ref{fig:hcop_outflow}, we chose to display 
emission originating only in those velocity ranges which best highlight 
the outflow emission.  Curiously, the SiO 2--1 emission in the region does 
not appear to trace the same outflow as seen in \hcop, or at least not the 
same part of it.  Figure~\ref{fig:sio_outflow} shows the integrated 
intensity of SiO 2--1 over two velocity ranges: one centered at the 
systematic velocity and the other exactly matching that used to highlight 
the red lobe of the \hcop\ outflow in Figure~\ref{fig:hcop_outflow}.  We 
did not detect any SiO 2--1 emission at the 3\sig\ level over the range of 
velocities which define the blue lobe of the \hcop\ outflow.  The only 
suggestion of an association between the \hcop\ outflow and the SiO 
emission is that they are both extended and collinear.  However, the 
velocity structures of the emission in the two molecules preclude their 
interpretation as parts of the same outflow.  A more likely scenario is 
that the SiO peak at the systematic velocity is associated with hot and/or 
shocked gas in the immediate vicinity of the massive protostar.  
Figure~\ref{fig:sio_outflow} shows that the SiO emission peak at the 
systematic velocity is spatially coincident with the H$_{2}$O maser 
identified by \citet{beu02b}.  The two red-shifted SiO emission peaks to 
the northwest are harder to explain.  These peaks may represent the red 
lobe of a separate outflow from the one delineated by the \hcop\ emission, 
or they may be entirely unrelated to outflow activity.  If these SiO 
emission peaks are 
unrelated to outflow emission--if, say, they were associated with other 
protostars--it is difficult to explain why they do not appear in other 
tracers.  If the SiO peaks are associated with outflow activity, we expect 
that this outflow would be either weaker or chemically very different from 
the prominent outflow seen clearly in \hcop.  

	The identity of the driving source of the \hcop\ outflow is not 
clear.  There is no candidate protostellar source at the point where the 
two outflow lobes overlap.  The driving source is probably the MSX point 
source, which is spatially coincident with the peak of MMS1 and the 3.6~cm 
continuum emission.  This source is presumably the principal massive 
protostar in the region.  Less clear, however, is whether the H$_{2}$O 
maser and SiO emission are also associated with this massive protostar, or 
whether they trace another source entirely.  Just as the MSX source aligns 
neatly with the continuum peak of MMS1, the H$_{2}$O maser aligns neatly 
with the peak of the systematic-velocity SiO emission.  Neither pair of 
features coincides with the other.  Hence, we speculate that there may be 
two massive protostars within the MMS1 core, one situated at the 
millimeter continuum peak and one displaced a few arcseconds to the south.

	There is no unambiguous evidence of an outflow emanating from 
MMS2.  The \hcop\ spectrum at the peak of MMS2 does show significant line 
wings (see Fig.~\ref{fig:infall_spectra}), but these could be caused by 
'spill-over' from the \hcop\ outflow emanating from the vicinity of MMS1.  
As can be seen in Figure~\ref{fig:hcop_outflow}, some of the \hcop\ 
emission which appears to be associated with the MMS1 outflow also 
overlaps MMS2.  There is some evidence in the \hcop\ data 
cube of a second \hcop\ outflow in the region, but the spatial resolution 
of the data make it difficult to associate it conclusively with either 
MMS1 or MMS2.  To determine conclusively whether there is an outflow 
emanating from MMS2, we would require data with higher spatial resolution 
and preferably in a transition which traces outflows more selectively 
than does \hcop.

\subsubsection{\meth}

\begin{figure} \begin{center} 
\includegraphics[width=0.8\columnwidth,angle=270]{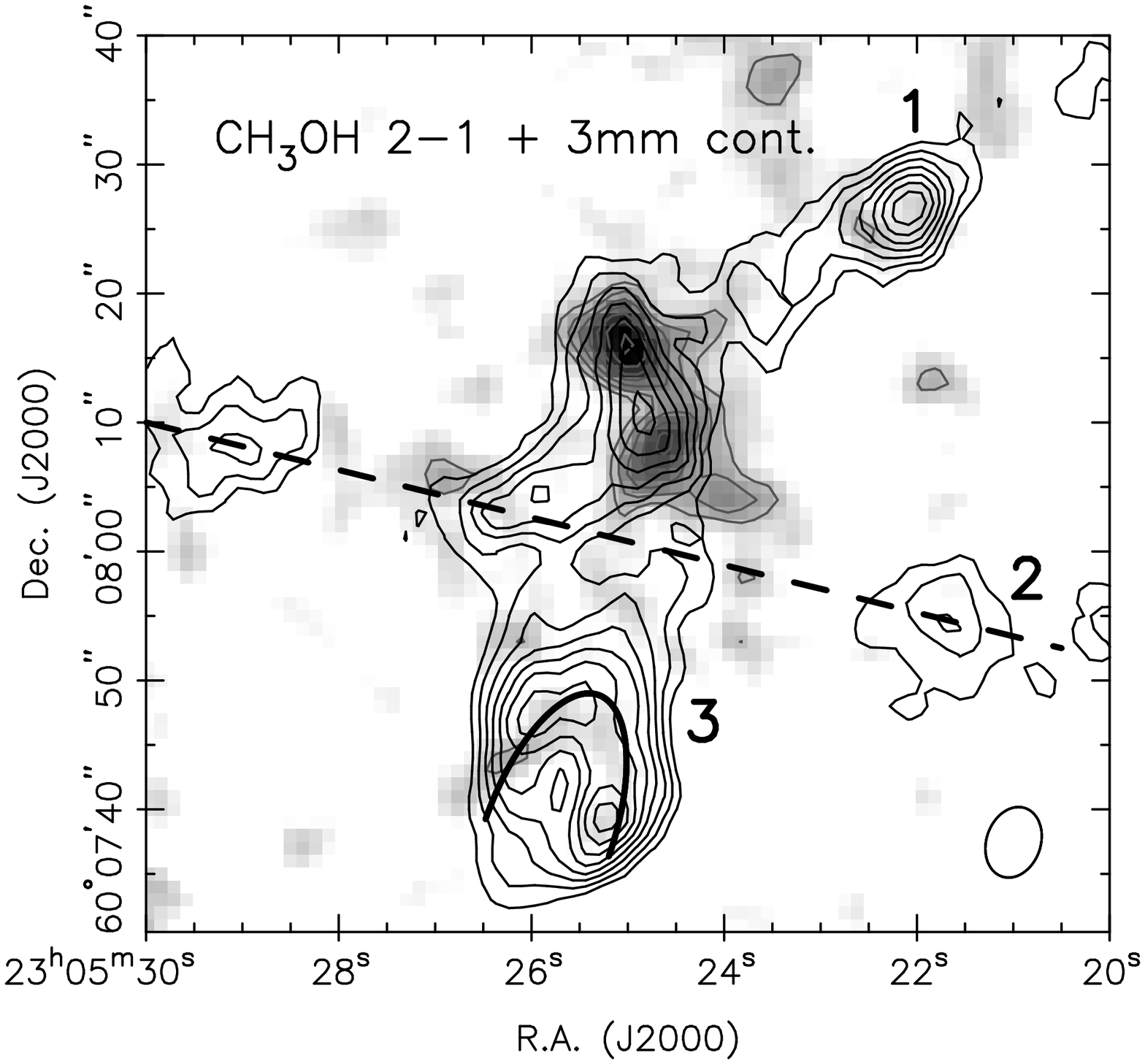} 
\caption{Integrated intensity of \meth\ 2--1 from -40.4 to -67.7 \kms, 
plotted over 3~mm continuum emission.  The black contours show the 
intensity of \meth\ 2--1 integrated over the two brightest components of 
the line, starting from 3\sig\ and increasing by increments of 2\sig, 
where \sig\ = 355 \mjbkms.  
The thin grey contours and greyscale represent the continuum as in 
Fig.~\ref{fig:contmodel}.  The size and orientation of the \meth\ 
synthesized beam are indicated by the ellipse in the lower right.  The 
features labeled in the diagram are (1) a possible \meth\ counterpart to 
the SiO outflow, (2) a possible second outflow seen only in \meth, and 
(3) a conical structure which may represent the edges of a conical cavity 
carved out by a third outflow.  Both of the outflows suggested by 
features (2) and (3) appear to lie approximately in the plane of the 
sky.\label{fig:meth_outflow}} \end{center} \end{figure}

As in many other high-mass star-forming regions, the \meth\ 2--1 emission 
in \iras\ is very complex.  Figure~\ref{fig:meth_outflow} shows the total 
integrated intensity of \meth\ in \iras.  The \meth\ emission in \iras\ is 
distributed widely in velocity and, in general, does not correlate neatly 
in either velocity or position with other outflow features.  The exception 
is the elongated feature labeled `1' in Figure~\ref{fig:meth_outflow}, 
which is coincident both spatially and in velocity with the SiO peaks 
shown in Figure~\ref{fig:sio_outflow}.  A prominent bar of \meth\ emission 
overlaps most, but not all, of the continuum-emitting region.  The 
emission peak of the \meth\ bar does not coincide with either continuum 
peak.

Although the \meth\ emission does not correlate well with the \hcop\ 
outflow, it can be interpreted as tracing several other possible outflows 
in the region.  For example, the dashed line labeled `2' in 
Figure~\ref{fig:meth_outflow} connects the \meth\ peak at the eastern edge 
of the image through a ridge of \meth\ emission to a second \meth\ peak 
near the western edge of the image.  In velocity, the \meth\ peaks at the 
two extremes of line `2' are separated by only a few \kms\, but their 
alignment suggests they may be part of an outflow lying nearly in the 
plane of the sky.  The lack of detected continuum emission at the position 
of either \meth\ peak supports this interpretation.  The driving source of 
this possible outflow is not evident.

	The most prominent \meth\ emission feature is the large parabolic 
structure, labeled `3', which dominates the southern half of the image.  
This feature has no counterpart in any other tracer we observed.  Its 
velocity structure suggests it may be the conical cavity of an outflow 
lying approximately in the plane of the sky.  Again, the source of this 
possible outflow is not evident, although the alignment of the feature is 
consistent with a driving source within the MMS2 core.  If the driving 
source did lie within MMS2, the second outflow lobe could be confused with 
the elongated bar of emission at the systematic velocity traced by the 
continuum emission and in several of the lines.

\subsection{A ``Flattened Rotating Object''}

	Four of the molecular transitions we observed have high critical 
densities and are known to trace dense star-forming gas.  These are the 
transitions for which integrated intensity maps are shown in 
Figure~\ref{fig:disklines}.  All four transitions trace similar structures 
in \iras: roughly elliptical distributions with long axes running from 
northwest to southeast, encompassing regions of high column density as 
traced by continuum emission (see Fig.~\ref{fig:contmodel}) The long axis 
of this elongated structure runs perpendicular to the axis of the 
prominent \hcop\ outflow.  A fit to the \htcop\ integrated intensity map 
shows that the position angle of the elongated feature is 
35$^{\circ}\pm$5$^{\circ}$ degrees, while that of the \hcop\ outflow is 
approximately 130$^{\circ}$, meaning that the position angles of the two 
features differ by about 95$^{\circ}$.

	As shown in Figure~\ref{fig:peaks}, the four dense gas tracers all 
peak at slightly different locations.  As we discussed in 
\S~\ref{sec:morph}, \ntwo\ and dust both trace the regions of highest 
column density.  Our \cs\ detection is too weak to determine the peak 
position accurately, except to say that it appears to correspond to the 
peak of the MMS1 core.  The peak of the \hcn\ emission matches that of the 
SiO emission and the position of the H$_{2}$O maser which, as we described 
in \S~\ref{sec:hcopsio}, is displaced by several arcseconds from the 
coincident dust continuum, radio continuum, and MSX peaks.  Notably, 
however, the maser and the emission peaks of \hcn\, SiO, and \cs\ all lie 
within one beam width of the MMS1 continuum peak, implying an association 
with MMS1.

\begin{figure*}
\begin{center}
\includegraphics[width=0.8\columnwidth,angle=270]{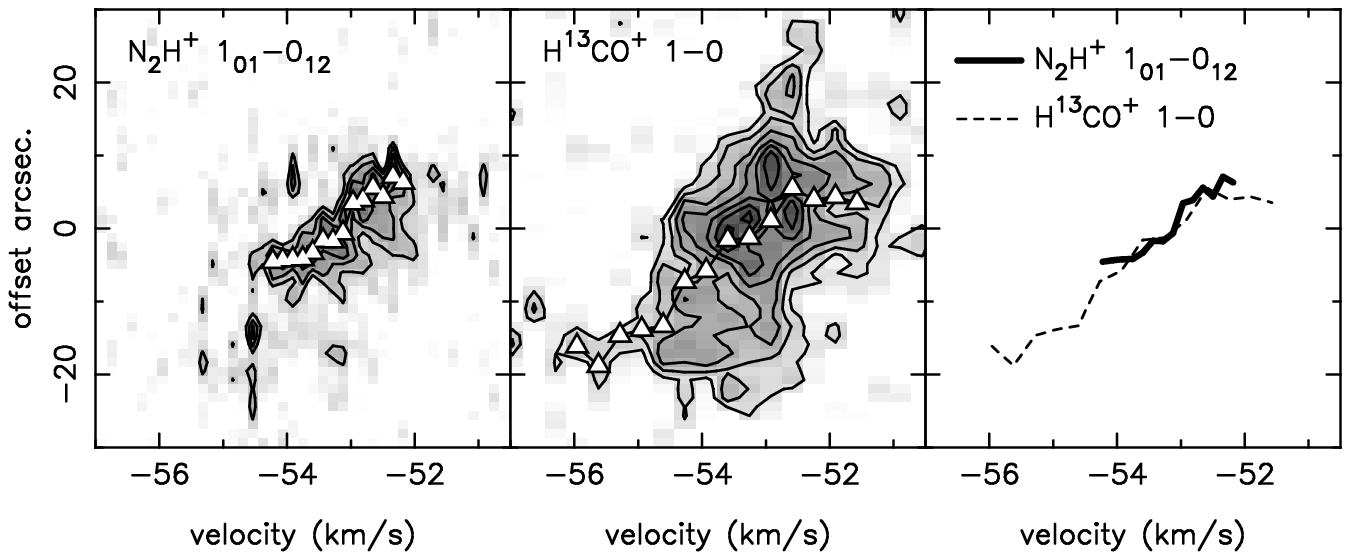}
\caption{Position-velocity slices taken along the long axis of the 
elongated structure seen in \ntwoi\ (\emph{left}) and \htcop\ 1--0 
(\emph{center}).  The position angle of the position-velocity cut is 
35$^{\circ}$.  Contours begin at 3\sig\ and increase by intervals of 
1\sig\ for \ntwoi\ and 2\sig\ for \htcop.  The white triangles in the 
left and center panels show the 
position-centroided velocities of the emission in each channel.  The right 
panel plots the two position-velocity curves on the same scale, 
illustrating the similarity between them.\label{fig:pvplot}}
\end{center} 
\end{figure*}

	The remaining dense gas tracer, \htcop, shows a distribution that 
requires some explanation.  \htcop\ peaks strongly at a point equidistant 
between the MMS1 and MMS2 continuum peaks, in a region of substantially 
lower column density as indicated by the dust emission.  The position of 
the \htcop\ peak does not match that of any other feature in the region 
(see Fig.~\ref{fig:peaks}).  It is unlikely that this peak is associated 
with any coherent physical structure; it is probably an artifact of 
depletion of \htcop\ onto dust grains at the column density peaks, 
leaving a remnant ``peak'' where depletion effects are least pronounced.  
Although \ntwo\ 1--0 and \htcop\ 1--0 peak at different locations, their 
kinematics clearly reveal that they trace the same contiguous structure in 
which the MMS1 and MMS2 peaks are embedded.  The axis ratio of the 
elongated structure seen in \htcop\ is 3.1.  Figure~\ref{fig:pvplot} shows 
position-velocity slices in \ntwoi\ and \htcop\ taken along the major axis 
of the elongated structure, at a position angle of 35$^{\circ}$.  As shown 
in the right panel of Figure~\ref{fig:pvplot}, both transitions exhibit 
the same velocity gradient, meaning that they trace the same large-scale 
structure.  Note that the signal-to-noise ratio of the \hcn\ and \cs\ data 
is too low to permit a meaningful analysis of their velocity structures.

	On large spatial scales, the elongated structure seen in 
Figure~\ref{fig:disklines} shares many characteristics in common with a 
stellar accretion disk.  For example, its long axis lies perpendicular to 
at least one outflow, there is a velocity gradient along its major axis, 
and it appears only in the continuum and emission from molecules which 
trace dense, non-outflowing gas.  However, the structure also has several 
characteristics which preclude its interpretation as an accretion disk, 
the first of which is its size.  The major axis of the \htcop\ emission 
over which there is a clear velocity gradient spans approximately 
25\arcsec, or about 87,000~AU at 3.5~kpc.  By comparison, protostellar 
accretion disks around low-mass stars typically have radii in the range 
50--500~AU \citep{dut96,kit02}.  Little is known about the plausible range 
of radii for accretion disks around massive protostars.  
Table~\ref{tab:bigdisks} gives an (incomplete) summary of the reported 
properties of candidate disks around massive protostars.  In most cases, 
the evidence that these objects are disks consists of three points: they 
are elongated in either the continuum or a dense gas tracer, the axis of 
elongation is approximately perpendicular to that of a detected outflow, 
and there is a velocity gradient along the long axis of the object 
observed in either molecular line or maser emission.  We note, however, 
that the range of reported physical properties of these disks is very 
large.  For example, their radii range over more than two orders of 
magnitude.  If we were to submit \iras\ as a candidate for this list based 
on its outflow properties and the shape and kinematics of its dense gas 
feature, it would only top out the list of an already very diverse sample.  
We note also that, while the molecular gas distribution in \htcop\ and 
possibly \ntwo\ could support the disk interpretation, the dust 
distribution does not.  The dust emission is better described as three 
strong peaks whereas an accretion disk should produce a centrally peaked 
distribution unless it had substantially fragmented.

\begin{deluxetable*}{lcccl}
\tablecaption{Candidate Massive Protostellar Disks\label{tab:bigdisks}}
\tablewidth{0pt}
\tablehead{\colhead{Object} & \colhead{Distance} & \colhead{Radius} & \colhead{Disk Mass} & \colhead{Ref.} \\ 
\colhead{} & \colhead{(kpc)} & \colhead{(AU)} & \colhead{(\msun)} & \colhead{} }
\startdata
G192.16-3.82 & 2 & 130 & 3 & \citet{shep01} \\
Cepheus A HW2 & 0.73 & 330 & 1--8 & \citet{patel05} \\
Becklin-Neugebauer Object & 0.46 & 800 & \nodata & \citet{jia05} \\
AFGL 5142 & 1.8 & 900 & \nodata & \citet{zha02} \\
IRAS 18089-1732 & 3.6 & $\sim$1000 & $\sim$16 & \citet{beu05} \\
NGC~7538 IRS 1N & 2.8 & 1000 & \nodata & \citet{pes04} \\
IRAS 20126+4104 & 1.7 & 1700 & 10 & \citet{ces97,ces99} \\
G24.78 A1 & 7.7 & 4000 & 23 & \citet{bel04} \\
G24.78 A2 & 7.7 & 4000 & 23 & \citet{bel04} \\
G24.78 C & 7.7 & 8000 & 23 & \citet{bel04} \\
G31.41 & 7.9 & 8000 & 23 & \citet{bel04} \\
NGC~7538~S & 2.8 & 15,000 & 400 & \citet{san03} \\
\enddata
\end{deluxetable*}

	How, then, should we interpret the kinematics of \iras\ and, more 
generally, of very large structures with rotation signatures surrounding 
massive protostars?  We suggest that the observations to date, including 
those summarized in Table~\ref{tab:bigdisks}, are consistent with the 
hypothesis that massive protostars may be fed by Keplerian accretion disks 
(e.g. IRAS 20126+4104), but that these accretion disks are smoothly 
connected to much larger, flattened, rotating structures.  These 
``flattened rotating objects'' are likely simply the remnants of the 
molecular cloud fragments from which the massive stars formed.  Existing 
observations, typically made with 
single interferometer configurations, only reveal rotational motions on 
the spatial scales visible with that interferometer configuration.  
Confirmation of the hypothesis that these rotating structures may extend 
from tens to thousands of AU would require multiple interferometric 
observations of each source with different interferometer configurations. 
 Such observations exist in only a few cases.  IRAS 20126+4104 is an 
instructive example: \citet{ces99} detect apparently Keplerian rotation on 
scales down to a few 100~AU using high-resolution CH$_{3}$CN observations 
while \citet{zha98} detect rotation out to radii of 5,000~AU in NH$_{3}$.  
We interpret our observations of \iras\ in a similar manner.  The 
optically thin \htcop\ line depicts a contiguous, flattened, 
slowly-rotating structure with a radius as large as 40,000~AU.  
\ntwo\ and dust continuum emission respectively reveal two and three 
condensations of higher column density within this apparently contiguous 
structure.  The cores detected in \ntwo\ and dust continuum emission may 
be fragmentation products of the original molecular cloud from which the 
\iras\ complex formed.

	It is also possible that the velocity gradient seen in \htcop\ and 
\ntwo\ may simply indicate that the elongated structure is a sheared 
filament.  This hypothesis cannot be readily ruled out using the data in 
hand.

\subsection{Infall}

\begin{figure}
\begin{center}
\includegraphics[width=1.3\columnwidth,angle=270]{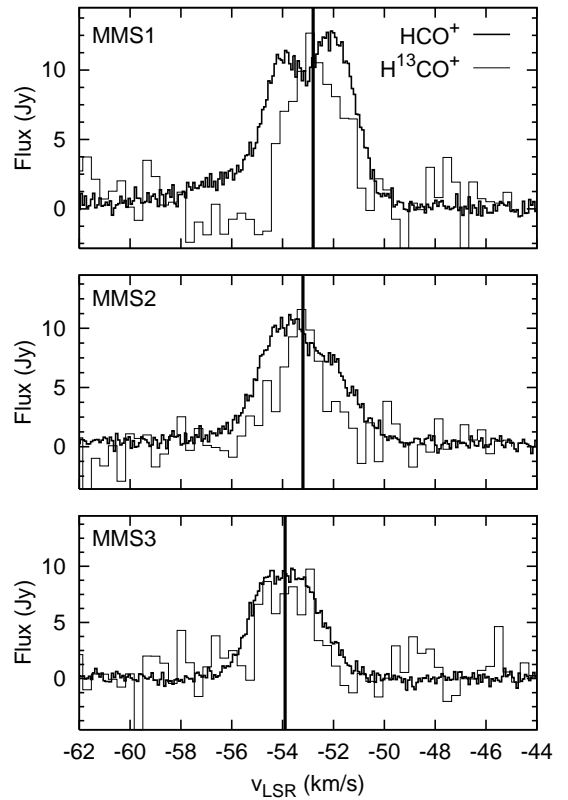}
\caption{Spectra of \hcop\ (\emph{thick black lines}) and \htcop\ 
(\emph{thin black 
lines}) at the positions of MMS1 (\emph{top}), MMS2 peak (\emph{middle}), 
and MMS3
(\emph{bottom}).  The vertical line in each plot indicates the fitted peak 
of the \htcop\ spectrum, an indication of the systematic velocity at that 
position.
\label{fig:infall_spectra}}
\end{center} 
\end{figure}

	\citet{zhou92} showed that, in a dense molecular cloud 
core undergoing some form of inside-out collapse, optically thick lines 
will be skewed to the blue by an amount which depends on the infall 
velocity.  The combination of a double-peaked, asymmetrically blue \hcop\ 
line whose self-absorption peak is centered at the velocity of the 
optically thin \htcop\ line has been interpreted as an indicator of infall 
in both low-mass \citep{gre00} and high-mass \citep{ful05} star-forming 
cores.  Figure~\ref{fig:infall_spectra} shows spectra of \hcop\ and 
\htcop\ at the three continuum peaks seen in Figure~\ref{fig:contmodel}.  
At the continuum peaks of both MMS1 and MMS2, the \hcop\ line appears 
self-absorbed, indicating that it is optically thick.  At the position of 
MMS1, the line is slightly skewed to the red.  We note, however, that 
there is significant outflow contamination in the spectrum of the core at 
this position.  At the position of MMS2, the line profile is skewed to the 
blue, suggesting it may be collapsing.

	Another possible interpretation of the double-peaked \hcop\ line 
profiles is that there are two sources along the line of sight.  This 
possibility would seem to be ruled out by the fact that the optically thin 
\htcop\ line is single-peaked throughout the region and it always peaks at 
the velocity of the \hcop\ self-absorption feature.  A third possible 
interpretation for the double-peaked and asymmetrically blue line profiles 
is that emission from the two cores is being blurred together in the 
comparatively large ($\sim10$\arcsec) \hcop\ beam.  To assess the 
likelihood of this possibility, we smoothed the \htcop\ data cube so that 
it had the same beam orientation and size as the \hcop\ beam.  Even in the 
smoothed data set, the \htcop\ line remains single-peaked at each 
continuum peak, thereby eliminating resolution effects as a likely 
explanation for the asymmetric line profiles.  Hence, we conclude that 
there is evidence for collapse in MMS2.

\section{Summary}
\label{sec:summary}

	We used millimeter interferometry to study the structure and 
kinematics of the massive star-forming region IRAS 23033+5951.  We find 
that the region resolves into at least three distinct cores, at least two 
of which have sufficient mass to form one or more massive stars.  There is 
strong evidence from the spatial coincidence of strong radio continuum 
emission, an MSX point source, and a 3~mm continuum peak that MMS1 is 
forming a single massive star.  This core also powers at least one outflow 
seen in \hcop\ 1--0 emission and possibly a second outflow seen in SiO 
2--1 and \meth\ 2--1.  The low virial ratio and moderately high density of 
MMS1 suggest it could be forming a massive star by collapse and disk 
accretion according to the ``monolithic collapse'' model.  The 
distribution of \ntwo\ in \iras\ strongly resembles that of the dust 
emission, suggesting that \ntwo\ is tracing the locations of highest 
column density.   Conversely, the fact that \htcop\ peaks between 
the column density peaks revealed by the dust and \ntwo\ emission probably 
reflects the depletion of \htcop\ onto grain mantles in the dense regions.  

	MMS2 appears to be quiescent: it harbors no maser or MSX sources 
and has no unambiguous outflow signature.  The spatial coincidence of its 
\ntwo\ and dust continuum emission peaks suggest that it is a cold core in 
an early stage of evolution.  With an estimated total mass (dust and gas) 
of 205~\msun, a virial ratio of 0.9, and a possible spectral signature of 
collapse, MMS2 appears poised to form a massive star of its own.  It may 
be an example of a ``Class 0'' massive protostar.

	Both of the cores described above are embedded in a smooth 
structure seen most clearly in the optically thin \htcop\ 1--0 line, but 
also seen in \hcn\ 1--0 and \hcop\ 1--0.  This structure has an axis ratio 
of 3.1, and its long axis lies almost exactly perpendicular to that of the 
\hcop\ outflow.  The \htcop\ emission peak of the structure is 
well-centered between the continuum peaks of the two bright cores.  There 
is a clear velocity gradient along the structure's long axis which is 
suggestive of rotation, although the size of the structure precludes its 
interpretation as a classical protostellar accretion disk.  We suggest it 
may either be the rotating, flattened remnant of the natal molecular cloud 
from which the two current cores formed, or simply a sheared filament.

\acknowledgements

	M.~A. Reid is currently supported by a fellowship from the 
Harvard-Smithsonian Center for Astrophysics but, at the time the data for 
this paper were acquired, he was supported by a scholarship from the 
Natural Sciences and Engineering Research Council of Canada.  B.~C. 
Matthews is supported by the National Research Council of Canada.  The 
BIMA array was operated with support from the National Science Foundation 
under grants AST 02-28963 to UC Berkeley, AST 02-28953 to U. Illinois, and 
AST 02-28974 to U. Maryland. FCRAO is supported by NSF grant AST 02-28993. 
 The radio continuum data shown in Figure~\ref{fig:peaks} were generously 
provided by H. Beuther.

\end{document}